\documentclass[3p,times,twocolumn]{elsarticle}
 \biboptions{comma,sort&compress}

\usepackage{graphicx}
\usepackage{amsmath}
\usepackage{ecrc}

\volume{00}

\firstpage{1}

\journalname{Nuclear and Particle Physics Proceedings}

\runauth{}

\jid{nppp}

\jnltitlelogo{Nuclear and Particle Physics Proceedings}

\usepackage{amssymb}

\usepackage[figuresright]{rotating}

\usepackage{mathrsfs} 
\usepackage{color}
\usepackage{slashed}
\usepackage{ulem}

\newcommand{\bean}{\begin{eqnarray*}}
\newcommand{\eean}{\end{eqnarray*}}

\newcommand{\gapproxeq}{\lower
.7ex\hbox{$\;\stackrel{\textstyle >}{\sim}\;$}}
\newcommand{\lapproxeq}{\lower
.7ex\hbox{$\;\stackrel{\textstyle <}{\sim}\;$}}

\newcommand\lsim{\mathrel{\rlap{\lower4pt\hbox{\hskip1pt$\sim$}}
    \raise1pt\hbox{$<$}}}
\newcommand\gsim{\mathrel{\rlap{\lower4pt\hbox{\hskip1pt$\sim$}}
    \raise1pt\hbox{$>$}}}
\newcommand{\ba}{\begin{array}}
\newcommand{\ea}{\end{array}}
\newcommand{\nn}{\nonumber}

\newcommand{\be}{\begin{equation}}
\newcommand{\ee}{\end{equation}}
\newcommand{\bear}{\begin{eqnarray}}
\newcommand{\eear}{\end{eqnarray}}

\newcommand{\ket}{\,\rangle}
\newcommand{\bra}{\langle \,}
\newcommand{\mL}{\mathscr{L}}

\newcommand{\mO}{\mathcal{O}}

\begin{document}

\begin{frontmatter}

\title{Heavy Resonances in the Electroweak Effective Lagrangian$^*$}
\cortext[cor0]{Talk given at 18th International Conference in Quantum Chromodynamics (QCD 15,  30th anniversary),  29 June - 3 July 2015, Montpellier - FR}

\author[label1]{J. Santos\fnref{fn1}}  
\address[label1]{Departament de F\'{\i}sica Te\`{o}rica, IFIC, Universitat de Val\`{e}ncia - CSIC, Apt. Correus 22085, 46071, Valencia, Spain}
\fntext[fn1]{Speaker.}
 \ead{Joaquin.Santos@ific.uv.es}

\author[label1]{A. Pich}
 \ead{pich@ific.uv.es}

\author[label2]{I. Rosell}
\address[label2]{Departamento de Ciencias F\'{\i}sicas, Matem\'{a}ticas y de la Computaci\'{o}n, Universidad CEU Cardenal Herrera\\ c/ Sant Bartomeu 55, 46115 Alfara del Patriarca, Valencia, Spain}
 \ead{rosell@uch.ceu.es}

\author[label3]{J.J. Sanz-Cillero}
\address[label3]{Departamento de F\'{\i}sica Te\'{o}rica and Instituto de F\'{\i}sica Te\'{o}rica, IFT-UAM/CSIC\\ Universidad Aut\'{o}noma de Madrid, Cantoblanco, 28049 Madrid, Spain}
 \ead{juanj.sanz@uam.es}

\pagestyle{myheadings}
\markright{ }
\begin{abstract}
As a first step towards the construction of a general electroweak effective Lagrangian incorporating heavy states, we present here a simplified version where only vector and axial-vector spin-1 triplets are involved. We adopt an effective field theory formalism,
implementing the electroweak chiral symmetry breaking $\mathrm{SU}(2)_L\times \mathrm{SU}(2)_R\to \mathrm{SU}(2)_{L+R}$, which couples the heavy states to the SM fields. At low energies, the heavy degrees of freedom are integrated out from the action and their effects are hidden in the low-energy couplings of the Electroweak Effective Theory, which can be tested experimentally. Short-distance constraints are also implemented, requiring a proper behaviour in the high-energy regime. We analyze the phenomenological constraints from the oblique parameters $S$ and $T$, at the next-to-leading order. Our results show that present data allow for strongly-coupled scenarios with massive bosons above the TeV scale.


\end{abstract}

\end{frontmatter}
\section{Introduction}
The LHC found a Higgs boson with the properties predicted by the Standard Model (SM). This discovery has completed the last piece of the puzzle, confirming the SM paradigm in particle physics. At the moment, there is no clear evidence of new physics below the TeV scale and the possibility of an energy gap gains importance. For this reason, effective field theories turn out to be a proper approach to search for new physics. The information on the high-energy degrees of freedom stays in the Low-Energy Constants (LECs) of the effective theory and they can be tested experimentally.

The Electroweak Effective Theory (EWET) provides a powerful framework to study many of the
open questions which remain unanswered within the SM.
In this direction, one obvious next step consists in adding new heavier states to the effective theory and investigate their possible signals at low energies.

In this article we study a simplified scenario with massive spin-1 triplets interacting with the bosonic sector of the electroweak theory. A more complete analysis can be found in \cite{project,Rosell:2015bra}. In section \ref{s2} we build the corresponding effective description which is then matched to the low-energy EWET in section \ref{s3}, and hence the LECs are determined. Another kind of conditions come from the ultraviolet (UV) completion of the effective theory. Thus, in section \ref{s4} some high-energy constraints must be imposed so that the theory is well-behaved. Phenomenology is analyzed in section \ref{s5}, where the oblique parameters $S$ and $T$ become a key factor in order to set bounds on the Resonance Theory parameters. Finally, some brief conclusions are given in section \ref{s6}.

\section{Building the Resonance Theory} \label{s2}
We call Resonance Theory to the effective field theory description which includes the SM particle content plus the additional massive states. 
We assume the SM pattern of Electroweak Symmetry Breaking (EWSB), so the theory is symmetric under $G\equiv \mathrm{SU}(2)_L \times \mathrm{SU}(2)_R$ and gets spontaneously broken to the custodial subgroup $H\equiv \mathrm{SU}(2)_{L+R}$. In this article, we perform a simplified analysis of the resonance theory:
\begin{itemize}
 \item Only the bosonic sector is studied. We consider the SM gauge bosons, the electroweak Goldstone bosons $\varphi^a$ and one Higgs-like scalar field $h$, with mass $m_h=125$ GeV, which is singlet under $H$.
 \item The resonance content is reduced to one vector and one axial-vector triplets, $V_{\mu\nu}$ and $A_{\mu\nu}$. We will use the antisymmetric formalism to describe these spin-1 fields \cite{rcht,RChTb,RChTc}.
 \item We assume that parity is a good symmetry of the strongly-coupled underlying theory.
\end{itemize}
According to these considerations, the lowest-order (LO) resonance interaction Lagrangian is
\begin{eqnarray}
 \mL &\!\! =&\!\! \frac{v^2}{4}\,\bra u_\mu u^\mu \ket \left( 1 + \frac{2 \,\kappa_W}{v}\,h\right) 
 \,+\,\frac{F_A}{2\sqrt{2}}\,\bra A_{\mu\nu} f_-^{\mu\nu}\ket
 \nn \\
 &\!\! +&\!\!  \frac{F_V}{2\sqrt{2}}\,\bra V_{\mu\nu} f_+^{\mu\nu}\ket\,+\,\frac{i G_V}{2\sqrt{2}}\,\bra V_{\mu\nu}[u^\mu,u^\nu]\ket\nn\\
 &\!\! +&\!\!  
 \sqrt{2}\, \lambda_1^{hA}\, \partial_\mu h\, \bra A^{\mu\nu} u_\nu \ket\,, 
 \label{res_lagr}
\end{eqnarray}
where the brackets stand for the $\mathrm{SU}(2)$ trace, and $u_\mu$, $f_{\pm\,\mu\nu}$ are chiral building blocks that involve Goldstone and gauge bosons, in agreement with the notation of Refs.~\cite{Pich:2013,Pich:2013fea}. The constant $\kappa_W$ measures the deviation from the SM in the Higgs coupling to the electroweak Goldstones.
A more complete analysis of the Resonance Theory, including the fermion sector, can be found in Ref.~\cite{project}.

\section{Constraining the EWET:
Determination of the LECs} 
\label{s3}

The EWET is the low-energy effective field theory with the same pattern of EWSB, but
with only the SM particle content. It can be obtained from the underlying Resonance Theory by integrating out the heavy fields from the action. The information on the high-energy degrees of freedom (the resonances) is encoded in the free parameters of the theory, the so-called LECs.

In order to estimate these parameters, we calculate the solution of the resonance equations of motion at LO in the momentum expansion, {\it i.e.}, chiral $\mO (p^2)$. Replacing the resonance fields in Eq.~(\ref{res_lagr}) by these solutions, we obtain an $\mO (p^4)$ EWET Lagrangian. As an example, we provide the resulting subset of low-energy operators for the parity even, purely bosonic sector without Higgs fields (the Longhitano's Lagrangian \cite{Longhitano:1980iz}):
\begin{eqnarray}
 \mL&\!\! \supset &\!\! \frac{1}{4}\, a_1\,\bra {f}_+^{\mu\nu} {f}_{+\, \mu\nu}^{\phantom{\mu}}- {f}_-^{\mu\nu} {f}_{-\, \mu\nu}^{\phantom{\mu}}\ket \nn \\
 &\!\! +&\!\! \frac{i}{2}\,(a_2-a_3)\, \bra {f}_+^{\mu\nu} [u_\mu, u_\nu] \ket \nn\\
 &\!\! +&\!\! a_4\,\bra u_\mu u_\nu\ket \, \bra u^\mu u^\nu\ket \,\,+\,\,  a_5\,\bra u_\mu u^\mu\ket^2 \nn\\
 &\!\! +&\!\! \frac{1}{2}\,H_1\, \bra {f}_+^{\mu\nu} {f}_{+\, \mu\nu}+ {f}_-^{\mu\nu} {f}_{-\, \mu\nu}\ket \,.
\end{eqnarray}
Once the resonances are integrated out, we obtain the following estimation for these LECs:
\begin{eqnarray}
 a_1 &\! =&\!  -\frac{F_V^2}{4M_V^2}\,+\,\frac{F_A^2}{4M_A^2}\,, \nn\\
 a_2-a_3 &\! =&\!  -\frac{F_V G_V}{2M_V^2}\,, \nn\\
 a_4 &\! =&\!  -a_5 \; =\; \frac{G_V^2}{4M_V^2}\,, \nn\\
 H_1 &\! =&\!  -\frac{F_V^2}{8M_V^2} \,-\, \frac{F_A^2}{8M_A^2} \,. \label{lecs}
\end{eqnarray}
The adopted procedure is completely analogous to the one developed in QCD to perform the matching between Chiral Perturbation Theory \cite{Weinberg:1979,chpt-SU2,Pich:1995bw,Ecker:1994gg} and Resonance Chiral Theory \cite{rcht,RChTb,RChTc,Pich:2002xy}.

\section{High-energy constraints} 
\label{s4}


Although the true fundamental electroweak theory remains still unknown, we can use the Resonance Theory as an effective framework which allows us to interpolate between the low-energy EWET description of Green functions and its assumed UV behaviour. We can then impose some properties that well-behaved high-energy theories must satisfy. As a consequence, new constrains arise.

\subsection{Form factors}
If we consider the interaction Lagrangian in Eq. (\ref{res_lagr}) and we calculate the two Goldstone boson vector form factor we obtain \cite{rcht,RChTb}:
\begin{align}
\bra \varphi^+(p_1)\, \varphi^-(p_2) |\, J^\mu_V\, | 0\ket \; &= \; (p_1-p_2)^\mu\; F_{\varphi\varphi}^V (s) \,,
 \nn\\[5pt]
 F_{\varphi\varphi}^V (s) \; = \; 1\,+\, &\frac{F_V G_V}{v^2}\,\frac{s}{M_V^2-s} \,.
\end{align}

A similar calculation of the scalar-Goldstone matrix element of the axial-vector current results in the axial form factor \cite{Pich:2013,Pich:2013fea}:
\begin{align}
\bra h(p_1)\, \varphi(p_2) |\, J^\mu_A\, &| 0\ket\; =\; (p_1-p_2)^\mu\; F_{h\,\varphi}^A (s) \,,
\nn\\[5pt]
F_{h\,\varphi}^A (s)\; =\;\kappa_W\, &\left( 1\,+\,\frac{F_A \lambda_1^{hA}}{\kappa_W v}\,\frac{s}{M_A^2-s} \right) \,.
\end{align}

Demanding the Resonance Theory to have a good UV completion, these two form factors should fall as $\mO (1/s)$ \cite{Brodsky-Lepage,Lepage:1980fj} and hence we get the conditions:
\begin{equation}
F_V\, G_V\; =\; v^2 \, ,\qquad\qquad F_A\, \lambda_1^{hA}\; =\; \kappa_W\, v \,. \label{ff}
\end{equation}

\subsection{Weinberg Sum Rules}

The two-point correlation function of a left and a right currents is an order parameter of (chiral) EWSB. In asymptotically-free gauge theories the difference $\Pi_{VV}(s)-\Pi_{AA}(s)$ vanishes at $s\to\infty$ as $1/s^3$ \cite{Bernard:1975cd}. This implies UV super-convergence properties, which have been largely used in QCD \cite{rcht,RChTb}, the so-called Weinberg Sum Rules (WSRs) \cite{WSR}. In the electroweak case, they constrain the gauge boson self-energies:
\begin{eqnarray}
 \frac{1}{\pi}\,\int_0^\infty ds\; \left[{\rm Im}\Pi_{VV}(s)-{\rm Im}\Pi_{AA}(s)\right] 
 &\! =&\! v^2\,,  
 \label{1wsr0}\\
 \frac{1}{\pi}\,\int_0^\infty ds\,s\;\left[{\rm Im}\Pi_{VV}(s)-{\rm Im}\Pi_{AA}(s)\right] &\! =&\! 0\,. 
 \label{2wsr0}
\end{eqnarray}
The first WSR in Eq.~(\ref{1wsr0}) is also valid in gauge theories with non-trivial UV fixed points \cite{Peskin:92}, but the second one could be questionable in some Conformal \cite{Orgogozo:2011kq} or Walking \cite{Appelquist:1998xf} Technicolour scenarios. For this reason, in the forthcoming sections we will consider separately the application of the 1st WSR and the two of them combined.

Calculating these two sum rules with our effective Lagrangian in Eq.~(\ref{res_lagr}), we obtain at LO:
\begin{eqnarray}
 F_V^2-F_A^2 &=& v^2\,,  \label{1wsr}\\
 F_V^2 M_V^2 - F_A^2 M_A^2 &=& 0 \, .\label{2wsr}
\end{eqnarray}
%
Note that these two conditions together imply $M_V<M_A$. At Next-to-Leading Order (NLO), {\it i.e.}, one loop, and taking into account the constraints in
Eq.~(\ref{ff}), the 2nd WSR gives the additional relation \cite{Pich:2013,Pich:2013fea}:
\begin{equation}
 \kappa_W\, =\, \frac{M_V^2}{M_A^2}\,.
\end{equation}
In this case, a big splitting of the resonance masses implies a big deviation from the SM in the Higgs coupling to $W^+W^-$ and $ZZ$ \ ($\kappa^{SM}_W=1$).

These high-energy constraints allow us to re-express the LECs in Eq.(\ref{lecs}) in terms of only the resonance masses:
\begin{eqnarray}
 a_1 &\!\! =&\!\!  -\frac{v^2}{4}\,\left( \frac{1}{M_V^2} + \frac{1}{M_A^2} \right) \,, \nn\\
 a_2-a_3 &\!\! =&\!\!  -\frac{v^2}{2 M_V^2} \,, \nn\\
 a_4 &\!\! =&\!\!  -a_5 \;\, =\;\, \frac{v^2}{4}\,\left( \frac{1}{M_V^2} - \frac{1}{M_A^2} \right) \,, \nn\\
 H_1 &\!\! =&\!\!  -\frac{v^2}{8} \, \left( \frac{1}{M_V^2} - \frac{1}{M_A^2} + \frac{2}{M_A^2-M_V^2} \right) \,. \quad\quad
\end{eqnarray}

\section{Phenomenology: Oblique Parameters} 
\label{s5}

Up to this point, we are dealing with seven parameters ($F_V$, $G_V$, $F_A$, $\lambda_1^{hA}$, $\kappa_W$, $M_V$, $M_A$) and five constraints. Hence, the parameter space has been remarkably reduced. The next step consists in relating these predictions with the data through the oblique parameters, which measure the possible presence of new-physics effects in the gauge boson self-energies.
We focus on the parameters $S$ and $T$ \cite{Peskin:92}. The first one parametrizes
new physics contributions to the difference between the $Z$ boson self-energy at $q^2=M_Z^2$ and $q^2=0$, while the parameter $T$ measures the breaking of custodial symmetry. A global fit to the electroweak data gives \cite{Baak:2014ora}:
\begin{equation}
 S\, =\, 0.05\pm0.11\, ,
 \qquad\quad 
 T= 0.09\pm0.13\, .
\end{equation}

In order to calculate these quantities we use the interaction Lagrangian (\ref{res_lagr}). At LO, only the $S$ parameter is non zero:
\begin{equation}
S_{LO}\, =\, 4\pi\,\left( \frac{F_V^2}{M_V^2}-\frac{F_A^2}{M_A^2} \right)\,, \qquad\; T_{LO}\, =\, 0\, .
\end{equation}
Applying the 1st WSR condition in Eq.~(\ref{1wsr}) and assuming $M_A>M_V$, this implies
a lower bound on the $S$ parameter \cite{Pich:2013,Pich:2013fea}:
\begin{equation}
 \frac{4\pi v^2}{M_V^2}\, <\, S_{LO}\, =\, 4\pi \left[ \frac{v^2}{M_V^2} + F_A^2\left( \frac{1}{M_V^2} - \frac{1}{M_A^2} \right)\right] .
\end{equation}
If we impose both the 1st and 2nd WSR conditions (\ref{1wsr}) and (\ref{2wsr}), an upper bound is obtained too \cite{Pich:2013,Pich:2013fea}:
\begin{equation}
 \frac{4\pi v^2}{M_V^2}\, <\, S_{LO}\, =\,\frac{4\pi v^2}{M_V^2} \left( 1+\frac{M_V^2}{M_A^2} \right)\, <\, \frac{8\pi v^2}{M_V^2} \,.
\end{equation}
Given these LO results, the resonance masses are bound to be\ $M_A>M_V>1.5$~TeV, and thus much heavier than the Higgs mass.

The NLO corrections can be computed by means of the dispersive representations \cite{Peskin:92,Pich:2013,Pich:2013fea}:
\begin{eqnarray}
 S &\!\! =&\!\! \frac{16\pi}{g^2 \tan \theta_W}\, \int_0^\infty \frac{{\rm d}s}{s}\; [\rho_S(s)-\rho_S(s)^{SM}]\,, 
 \nn\\[5pt]
 T &\!\! =&\!\! \frac{4\pi}{g^2 \cos^2 \theta_W}\, \int_0^\infty \frac{{\rm d}s}{s^2}\; [\rho_T(s)-\rho_T(s)^{SM}]\,, 
\end{eqnarray}
being $\rho_S(s)$ the spectral function of the $W_3B$ correlator and $\rho_T(s)$ the spectral function of the difference between the self-energies of the neutral and charged Goldstone bosons. We work at LO in the gauge couplings $g$ and $g'$ and only the lightest two-particle cuts have been considered, since higher-energy channels with massive resonances are very suppressed \cite{Pich:2012}.

\begin{figure}[t]
 \includegraphics[width=0.45\textwidth]{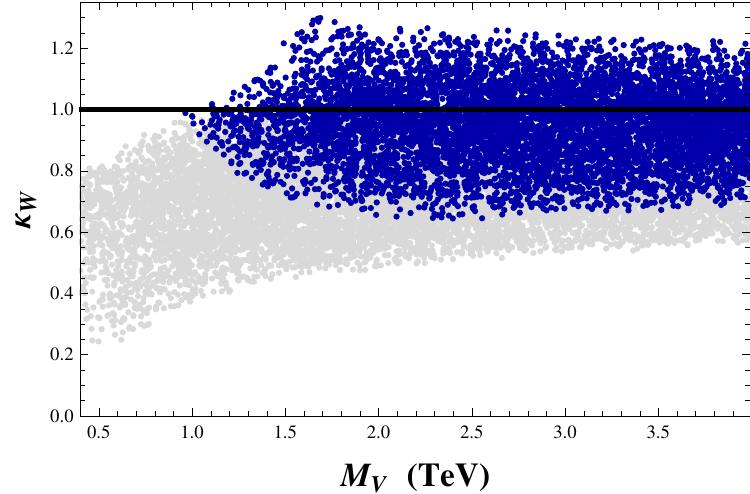}
 \caption{Allowed 68\% CL region when only the 1st WSR is imposed. The grey and blue areas
 correspond to large ($0.02<M_V/M_A<0.2$) and small ($0.2<M_V/M_A<1$) splittings, respectively \cite{Pich:2013,Pich:2013fea}.}
 \label{mvkw}
\end{figure}

If only the 1st WSR is imposed, the oblique parameters at NLO turn out to be~\cite{Pich:2013,Pich:2013fea}:
\begin{eqnarray}
S_{NLO} &\!\!\geq&\!\! 
 \frac{4\pi v^2}{M_V^2} \,+\, \frac{1}{12\pi}\left[ \log \frac{M_V^2}{m_h^2} \,-\, \frac{11}{6} \right.
 \nn\\
  &\!\! -&\!\! \left. \kappa_W^2\left( \log\frac{M_A^2}{m_h^2} \,-\, \frac{17}{6} \,+\, \frac{M_A^2}{M_V^2}\right)\right]
\nn\\  &\!\! +&\!\! 
\mO\left( \frac{m_h^2}{M_{V,A}^2}\right) \,,  
\\[5pt]
 T_{NLO}&\!\! =&\!\!\frac{3}{16\pi\cos^2\theta_W}\left[1 \,+\, \log\frac{m_h^2}{M_V^2}  \right. \nn\\
  &\!\! -&\!\! \left.\kappa_W^2\left(1\,+\,\log\frac{m_h^2}{M_A^2}\right)\right]\,+\,\mO\left( \frac{m_h^2}{M_{V,A}^2}\right)\,.
\end{eqnarray}
For the parameter $S$ just a lower bound in terms of the resonance masses and $\kappa_W$ can be set. Again, $M_V<M_A$ is assumed. Figure \ref{mvkw} represents the allowed 68\% CL region in the plane $(\kappa_W, M_V)$, varying $M_V/M_A$ between 0.02 and 1. Values of the Higgs gauge coupling $\kappa_W$ very different from the SM can only be obtained with a large splitting of the vector and axial-vector masses. For $0.5<M_V/M_A<1$, 
masses above 1.5 TeV are required and $\kappa_W>0.84$ at $68\%$ CL.

\begin{figure}
 \includegraphics[width=0.45\textwidth]{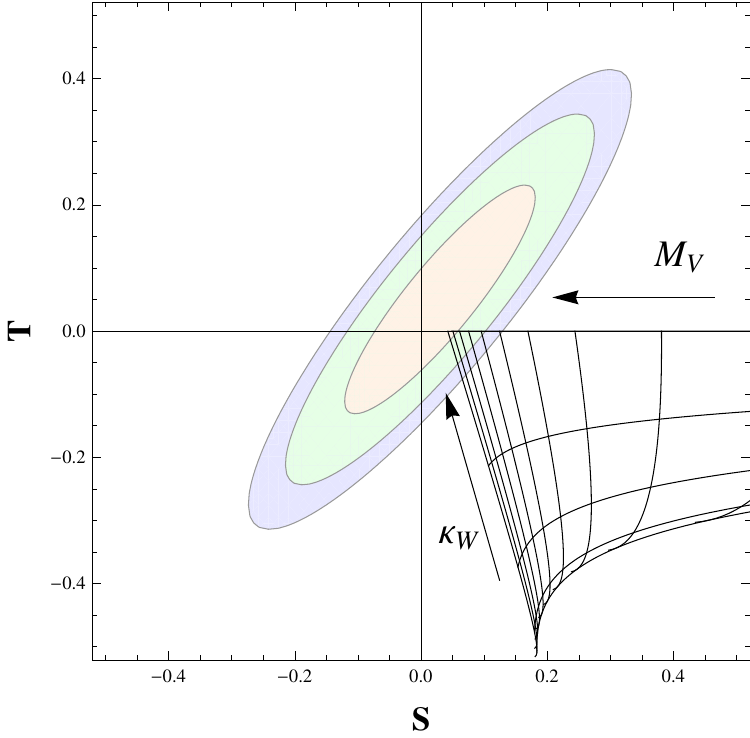}
 \caption{Allowed 68\%, 95\% and 99\% C.L. regions for the oblique parameters $S$ and $T$, with both WSRs assumed. The grid shows different values for the parameters $\kappa_W$ ($0.00, 0.25, 0.50, 0.75, 1.00$) and $M_V$ (from 1.5 to 6.0 TeV, at intervals of 0.5 TeV). The arrows indicate the growing directions of these parameters \cite{Pich:2013,Pich:2013fea}.}
 \label{st}
\end{figure}

Combining both WSRs at NLO, we can express the oblique parameters in terms of just the resonance masses:
\begin{eqnarray}
S_{NLO}&\!\! =&\!\! 4\pi v^2 \left( \frac{1}{M_V^2}\,+\, \frac{1}{M_A^2}\right) 
\nn\\
 &\!\! +&\!\! \frac{1}{12\pi} \left[ \log\frac{M_V^2}{m_h^2} - \frac{11}{6} + \frac{M_V^2}{M_A^2} \log\frac{M_A^2}{M_V^2}\right.
 \nn\\
 &\!\! -&\!\! \left. \frac{M_V^4}{M_A^4}\left(\log\frac{M_A^2}{m_h^2} - \frac{11}{6}\right)\right]\,+\,\mO\left( \frac{m_h^2}{M_{V,A}^2}\right)\,, 
\\[5pt]
T_{NLO} &\!\! =&\!\!\frac{3}{16\pi\cos^2\theta_W}\left[1 \,+\, \log\frac{m_h^2}{M_V^2}  \right. 
\nn\\
&\!\! -&\!\!  \left.\frac{M_V^2}{M_A^2}\left(1\,+\,\log\frac{m_h^2}{M_A^2}\right)\right]\,+\,\mO\left( \frac{m_h^2}{M_{V,A}^2}\right)\,.
\end{eqnarray}
Figure \ref{st} shows different CL regions of the $S$ and $T$ parameters. We can see how values of $\kappa_W$ close to 1 fit inside the ellipses and heavier resonance masses are preferred. We obtain $\kappa_W\in[0.97,1]$,  $M_V>5$ TeV at 68\% CL, and $\kappa_W\in[0.94,1]$,  $M_V>4$ TeV at 95\% CL. These results agree with the LHC findings and set more stringent bounds with only one mild hypothesis: the validity of the second WSR.

\section{Conclusions} \label{s6}

Strongly-coupled scenarios are still a great open possibility for new physics 
beyond the SM. Using an effective field theory approach which implements the known pattern of EWSB, we have analyzed generic couplings of spin-1 massive states to the SM Higgs, electroweak Goldstones and gauge bosons. A much more general situation is considered in Ref.~\cite{project}.

Integrating out the massive resonances, we have identified their effects on the LECs of the EWET. If future LHC data finds out a non-zero contribution in any of those LECs, one could then infer information about the hypothetical origin of that effect.

Studying the short-distance properties of Green function in the Resonance Theory and requiring a proper UV behaviour, one finds constraints on the resonance couplings which in turn allow for much sharper predictions on the EWET LECs.

We have also performed a one-loop calculation of the oblique parameters $S$ and $T$ within this framework. The results are compatible with the current experimental data and pin down  the gauge coupling of the Higgs-like boson to be close to the SM predictions, with better accuracy than the direct LHC measurements. Imposing only the 1st WSR, which is expected to be valid in all reasonable dynamical scenarios, our analysis shows that the vector and axial-vector states should be heavier than 1.5 TeV to comply with the electroweak precision data. Stronger bounds arise when the 2nd WSR is assumed too; one finds then that $M_V<M_A$, and a small splitting between these two masses is favoured by the experimental data.

\section{Acknowledgements}
We want to thank the organizers for their effort to make this conference such a successful event. We also want to thank the support by the Spanish Government and ERDF funds from the European Commission [FPA2011-23778, FPA2013-44773-P and FPA2014-53631-C2-1-P], by the Spanish {\it Centro de Excelencia Severo Ochoa} Programme [Grants SEV-2012-0249 and SEV-2014-0398], the Generalitat Valenciana [PrometeoII/2013/007] and La Caixa [PhD grant for Spanish universities].

\end{document}